# Metallic delafossite thin films for unique device applications

Takayuki Harada[1,2,a] and Yoshinori Okada[3]

[1] MANA, National Institute for Materials Science, 1-1 Namiki, Tsukuba, Ibaraki 305-0044, Japan

[2] JST, PRESTO, 4-1-8 Honcho, Kawaguchi, Saitama, 332-0012, Japan

[3] Quantum Materials Science Unit, Okinawa Institute of Science and Technology (OIST), Okinawa 904-0495, Japan

[a] Correspondence to: HARADA.Takayuki@nims.go.jp

Metallic delafossites ($AB$O$_2$) are layered oxides with quasi-two dimensional conduction layers. Metallic delafossites are among the most conducting materials with the in-plane conductivity comparable with that of elemental metals. In this perspective, we will discuss basic properties and future research prospects of metallic delafossites, mainly focusing on thin films and heterostructures. We exemplify the fascinating properties of these compounds such as high conductivity and surface polarity and discuss how it can be utilized in thin films and heterostructures.

## 1. INTRODUCTION

Research on quantum materials has deepened our knowledge of condensed matter physics and expanded the choice of materials for future thin-film devices. Metallic delafossites PdCoO$_2$, PdCrO$_2$, PdRhO$_2$, and PtCoO$_2$, a family of materials that will be the focus of this perspective, are natural superlattices of two-dimensional (2D) conducting sheets and correlated insulating layers.[1-3] In the rigid oxide framework shown in Figure 1, metallic delafossites demonstrate intriguing physical properties driven by high electron mobility and surface polarity. These compounds are also chemically inert and thermally stable, with a decomposition temperature of ~800°C or higher in air.[1] The high stability of metallic delafossites give them great potential for use in thin-film devices.

After an overview of the basic properties of metallic delafossites (section 2), this paper will focus on two important characteristics of these materials with respect to device applications: high mobility/conductivity, which give rise to intriguing transport phenomena (section 3 and 4), and surface polarity, which leads to a large work function and emergent spin polarized surface states (section 5 and 6). While high mobility/conductivity and surface polarity are essentially independent properties, metallic delafossites in characteristic layered crystal structures demonstrate these properties. The physical properties of metallic delafossites have mainly been revealed through the study of bulk single crystals[4], with limited research on thin films. For metallic delafossites to be used for device applications, it will be crucial to accelerate research on thin films, which includes making heterostructures by design (section 6). For this purpose, as described in section 7, we emphasize the importance of direct imaging of metallic delafossite thin films using spectroscopic imaging scanning tunneling microscopy (SI-STM).

## 2. PHYSICAL PROPERTIES OF METALLIC DELAFOSSITES

In this section, a brief overview is given on the physical properties of metallic delafossites. For details, the reader can refer to existing reviews on single crystals,[4,5] thin films,[6] and theory.[7] As shown in Figure 1(a), the crystal structure of metallic delafossites consists of ionic layers of $A^+$ and $[B$O$_2]^-$ octahedra. At present, five compounds in the delafossite family have been shown to exhibit metallic

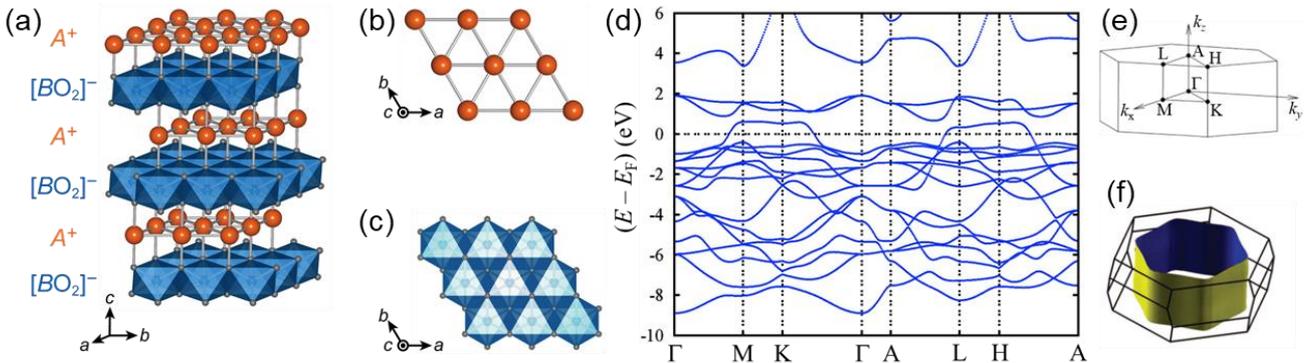

**FIG. 1.** (a) The crystal structure of metallic delafossites $AB$O$_2$. The $A^+$ and $[B$O$_2]^-$ layers are noted. (b) The crystal structures of $A^+$ layers and (c) the $[B$O$_2]^-$ layers. (d) The electronic band structure, (e) the Brillouin zone of the hexagonal unit cell, and (f) the Fermi surface of PdCoO$_2$ calculated by density-functional theory (DFT) in the generalized gradient approximation (GGA).[79] Reprinted with permission from ACS (Copyright 2008, American Chemical Society)

electrical conduction: $PdCoO_2$, $PdCrO_2$, $PdRhO_2$, $PtCoO_2$, and $AgNiO_2$.[4] Here, we will focus on the first four compounds, in which $A^+$ ($Pd^+$ or $Pt^+$) ions mediate electrical conduction, and will refer to them collectively as metallic delafossites. Hereafter, we will not discuss $AgNiO_2$ as in this compound, the $B$-site Ni ions contribute to electrical conduction, resulting in considerably different physical properties from the other four compounds.[8-10]

In each layer, $A^+$ ions or $BO_6$ octahedra form 2D triangular lattices (Fig. 1(a)–(c)). The $A^+$ layers are 2D conductive sheets while the $[BO_2]^-$ layers are insulating in nature. Because of this layered crystal structure, metallic delafossites show quasi-2D electrical conduction. Conductivity in the ab-plane ($\sigma_{ab}$) of metallic delafossites is generally orders of magnitude higher than along the c-axis ($\sigma_c$). In the case of $PdCoO_2$ at room temperature, $\sigma_{ab}$ ~ $3.8 \times 10^7$ S/m while $\sigma_c$ ~ $9.3 \times 10^4$ S/m. At cryogenic temperatures, the conductivity of $PdCoO_2$ increases by a factor of ~400 for $\sigma_{ab}$,[11] resulting in an electron mean free path length $l$ ~ 21.4 μm.[4, 12]

As will be discussed in section 3, the quasi-2D high-mobility electrons generate intriguing electrical transport phenomena as recently shown in high-purity single crystals.[13-18] The band structure and Brillouin zone of $PdCoO_2$ is shown in Figures 1(d) and (e). The dispersive band that crosses the Fermi level is predominantly attributed to Pd 4d and Pd 5s states.[19] These electronic states form the nearly cylindrical Fermi surface shown in Figure 1(f), reflecting the quasi-2D electrons in the Pd layers.

Regarding magnetic properties, $PdCrO_2$ shows antiferromagnetic transition with the Néel temperature $T_N$ ~ 37.5 K while the bulk states of $PdCoO_2$, $PdRhO_2$, and $PtCoO_2$ are nonmagnetic.[4] In $PdCrO_2$, the $Cr^{3+}$ ions have localized spins with $S = 3/2$.[20, 21] As shown in Figure 1(c), the $CrO_6$ octahedra form geometrically frustrated 2D spin triangular lattices, which influence the high-mobility electrons in the Pd sheets and cause an unconventional anomalous Hall effect.[20] Although $PdCoO_2$, $PdRhO_2$, and $PtCoO_2$ are nonmagnetic in bulk because of the fully occupied $t_{2g}$-like orbitals, recent experiments have indicated the existence of spin-dependent electronic states at the surface, which will be discussed further in sections 5–7.[22-24]

## 3. HIGH MOBILITY AND HIGH DENSITY QUASI-2D ELECTRONS

As described in section 1, one of the most remarkable characteristics of metallic delafossites are their high electrical conductivity. Figure 2 compares the metallic delafossites and various high-mobility 2D electronic systems: 2D materials, quasi-2D materials, and 2D electrons at surfaces/interfaces (Figs 2(a)-(c)). As shown in Fig. 2(d), currently known high mobility electronic systems have relatively low carrier densities of $n_{2D} < 10^{14}$ $cm^{-2}$. In contrast, metallic delafossites, plotted in red in Fig. 2(d), are high mobility quasi-2D systems with comparably high carrier densities of $n_{2D}$ ~ $10^{15}$ $cm^{-2}$. Thus, metallic delafossites are unique and ideal for exploring mesoscopic electrical transport phenomena in high mobility, high carrier density materials (Fig. 2(d)).

To compare the cleanness of 2D conduction channels, mean free path $l$ is plotted in Fig. 2(e) for the same set of data as Fig. 2(d). The mean free path of metallic delafossites compares well with most of high-mobility materials, showing exceptional cleanness of the conduction channels of metallic delafossites (Fig. 2(e)). As demonstrated by $PdCoO_2$ single crystals at cryogenic temperatures, the 2D electrons in Pd sheets have impressively long electron scattering lengths, with an electron mean free path length $l$ ~ 21.4 μm[12] and a phase coherence length $l_\phi > 10$ μm.[17] An interesting consequent phenomenon observed for the high-mobility electrons in $PdCoO_2$ is hydrodynamic electron flow.[13] The flow of electrons in solids is normally far different from the hydrodynamic flow of an ordinary fluid like water. In ordinary conductors, electrons are scattered by phonons and impurities, resulting in momentum relaxation and the generation of resistance. On the other hand, if the total momentum of electrons is conserved inside a conductor and is predominantly relaxed at the edges, electrical transport resembles the hydrodynamic motion of a fluid. To achieve this, the rate of momentum-relaxing scattering inside a conductor (electron-impurity, electron-phonon, and Umklapp scattering) should be minimized. In this situation, the channel resistance is partly given by the viscosity of electron fluid. The viscosity of electron fluid is governed by the rate of momentum-conserving electron-electron scattering. This was recently demonstrated by Moll and colleagues who fabricated microchannels in $PdCoO_2$ single crystals using a focused ion beam (FIB) technique.[13] The low rate of momentum-relaxing scattering in ultrapure $PdCoO_2$ single crystals results in electrical transport through these microchannels that can be described as hydrodynamic motion.[13, 25] Furthermore, various mesoscopic high mobility electrical transport phenomena, such as ballistic[16, 18, 26] or phase-coherent transport[17], have been demonstrated in high-purity single crystals of $PdCoO_2$.

Along with the above-mentioned transport phenomena, the high carrier density in metallic delafossites is predicted to cause additional physical effects. Because of the high carrier density, the Fermi surfaces of metallic delafossites are considerably larger than other quasi-2D systems, such as semiconductor heterostructures (Fig. 1(f)). This can cause the many-body effects, such as quasiparticle formation and/or an anisotropic Fermi surface shape, to appear with the high mobility transport phenomena. For example, because of the large hexagonal cylindrical Fermi surface, the trajectory of ballistic transport is not isotropic as in ordinary 2D systems but highly directional and dominantly oriented in six directions (Fig. 3(a), (b)).[16, 18, 26] Identifying exotic consequences of many-body effects in mesoscopic electrical transport is an intriguing challenge, which may be facilitated by real-space probing techniques (section 7).

While mesoscopic electrical transport phenomena in metallic delafossites have been mainly studied in FIB-patterned bulk single crystals, these fascinating transport properties are motivating thin-film researchers to grow high-purity $PdCoO_2$ in order to make mesoscopic devices using conventional lithography techniques. Once realized, exotic transport in thin films could be expanded to the study of spin transport or the superconducting proximity effect by making thin-film heterostructures, as will be discussed in section 6 and 7. Signatures of phase-coherent transport in $PdCoO_2$ thin films have already been reported in mesoscopic structures.[27] Although the $l_\phi$ and $l$ of thin-film mesoscopic structures are currently $l_\phi$ ~ 100 nm and $l$ ~ 10 nm,[27] improving thin film quality could extend these scattering lengths.

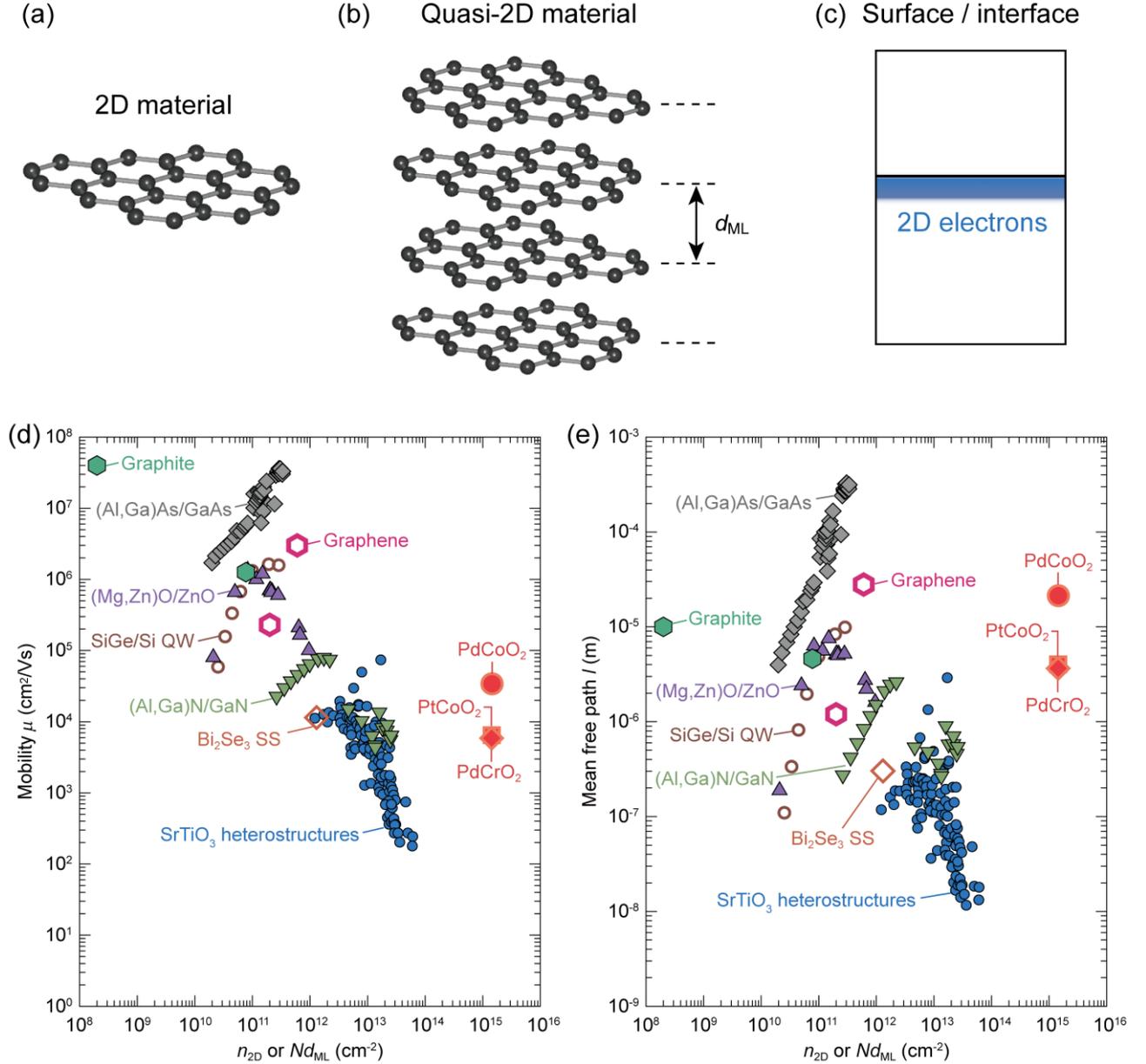

**FIG. 2.** Comparison of various (quasi-)2D high mobility electronic systems. The schematics of (a) 2D material, (b) quasi-2D material, and (c) 2D electrons at a surface/interface. (d) Charge carrier mobility $\mu$ of selected materials. 2D materials: suspended graphene at ~ 5 K[80] and graphene on h-BN below 130 K (open magenta hexagons).[80, 81] Quasi-2D materials: bulk single crystals of $PdCoO_2$ (red circle), $PtCoO_2$ (red square), and $PdCrO_2$ (red diamond) at 4 K[12] and graphite (green hexagon) at 4.2 K[82, 83] or 3 K.[84] Semiconductor interfaces: (Al,Ga)As/GaAs heterostructures (black diamonds),[85, 86] SiGe/Si quantum wells (QW, brown open circles) at 0.3 K,[87] (Mg,Zn)O/ZnO heterostructures (purple up-pointing triangles) at 500 mK or ≤ 100 mK,[88] (Al,Ga)N/GaN heterostructures (green down-pointing triangles) at 20 K[89] or 0.3 K,[90] and $SrTiO_3$-based heterostructures (blue circles) at low temperature.[91] Surfaces of topological insulators: the surface states (SS) of $Bi_2Se_3$ at 1.5 K.[92] For the quasi-2D systems, the sheet carrier density for a single conductive layer is calculated as $n_{2D} = N d_{ML}$ using the carrier concentration $N$ and the interlayer distance $d_{ML}$ schematically shown in (b). The carrier mobility of $PdCoO_2$, $PtCoO_2$, and $PdCrO_2$ are calculated as $\mu = 1/eN\rho$, where $e$ is elemental charge and $\rho$ is resistivity. The reported $ab$-plane resistivity is used for $\rho$. (e) Mean free path $l$ of charge carriers in the same materials as (d). For graphene and graphite, we used the reported values of $l$ in the literature.[80-82, 84] For the other materials, the mean free path $l$ is calculated using the formulas for circular Fermi surfaces around valleys: $l = \frac{\hbar k_F \mu}{e}$ and $k_F = \left(\frac{4\pi n_{2D}}{g_s g_v}\right)^{\frac{1}{2}}$, where $\hbar$ is the reduced Planck constant, $k_F$ is the Fermi wave vector, $g_s$ is the spin degeneracy, and $g_v$ is the valley degeneracy. For $SrTiO_3$ heterostructures, the non-circular 2D Fermi surface could result in deviation from the calculated $l$. We also note that the mean free path is proposed to be limited by the sample size for the graphene data.[80, 81]

## 4. TOWARD THIN FILMS WITH HIGH ELECTRON MOBILITY

In this section, we will discuss the possibility of achieving high electron mobility metallic delafossite thin films. Metallic delafossite thin films were first fabricated in the 1980s by annealing amorphous precursors deposited by sputtering.[28] Recently, several groups have reported the fabrication of c-axis oriented thin films of PdCoO$_2$,[29-33] PdCrO$_2$,[33-35] and PtCoO$_2$[33] by pulsed laser deposition (PLD),[29, 31, 34] molecular beam epitaxy,[30, 32] and solution-based processes.[33]

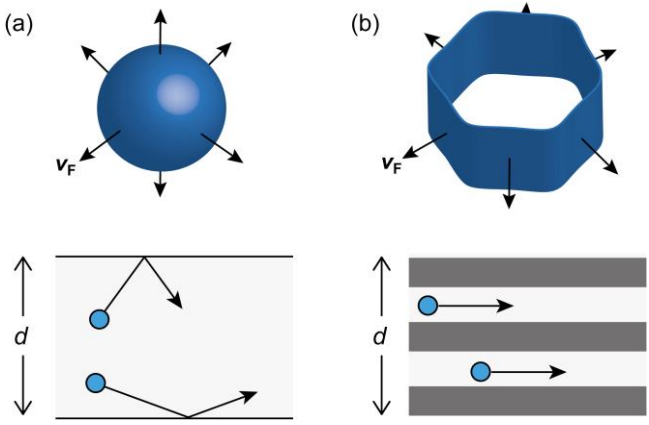

**FIG. 3.** (a) A spherical Fermi surface (top) and a schematic of electrons travelling with $v_F$ in a thin film with thickness $d$ (bottom). (b) A schematically drawn hexagonal Fermi surface of PdCoO$_2$ (top) and electrons travelling with $v_F$ along the conductive layer of Pd (bottom). The total thickness of the thin film is noted as $d$. We note that the warping in the PdCoO$_2$ Fermi surface results in finite component of $v_F$ perpendicular to the conductive layer.

Is it possible to increase the electron mobility of thin films to the level of single crystals? With the exception of metallic delafossites, studies on the resistivity of thin metal films have a long history. Studies on elemental metals such as Au,[36] Pt,[37] Co,[38] and Cu[39, 40] have shown that the resistivity of metals scales with thin film thickness and generally increases as thickness is reduced down to a few nanometers, which can be explained by the surface scattering of electrons. According to theory, the resistivity, $\rho$, of three-dimensional metals is expected to increase as the thickness, $d$, decreases, with $\rho \propto 1/[d\ln(l_{bulk}/d)]$,[41, 42] $\rho \propto 1/d$,[40] or $\rho \propto 1/d^2$,[43-45] depending on the model, where $l_{bulk}$ is the bulk mean free path.

Because of the quasi-2D nature of PdCoO$_2$, a reduction in thickness may not cause a serious surface scattering effect since electron mobility could be maintained, as seen in other quasi-2D systems like graphite. The directional transport effect found in single crystals of PdCoO$_2$ is an encouraging result that supports this hypothesis. According to transport measurements of PdCoO$_2$ single crystals,[16, 18, 26] the ballistic transport of electrons is highly anisotropic, reflecting the hexagonal cylindrical shape of the Fermi surface. The spherical Fermi surfaces of alkali metals are compared with those of PdCoO$_2$ in Figure 3(a) and (b).

The velocity of electrons, $v_n$, is expressed as,

$$v_n(k) = \frac{1}{\hbar} \nabla_k \varepsilon_n(k)$$

, where $n$ is the band index, $k$ is the wavevector, $\hbar$ is the reduced Planck constant, and $\varepsilon_n(k)$ is the energy of an electron with $k$ in the $n$-th band.[46] The Fermi velocity, $v_F$, is therefore perpendicular to the Fermi surface and is isotropic in alkali metals that has additional one electron to the closed-shell configuration. In contrast, the $v_F$ of PdCoO$_2$ is dominantly oriented in six directions in the ab-plane (Fig. 3, top). Because of the directional ballistic transport, c-axis oriented thin films of metallic delafossites may not be significantly affected by surface scattering (Fig. 3, bottom). Although the warping in the Fermi surface of PdCoO$_2$ causes a finite component of $v_F$ and sets a limitation to the above discussion,[47] the nearly cylindrical Fermi surface of PdCoO$_2$ should be a predominant advantage for achieving long mean free path in thin films.

Table 1 summarizes the challenges toward high-mobility thin films of metallic delafossites and reported approaches. So far, PdCoO$_2$ is most studied for thin films among the metallic delafossites. The typical growth condition of PdCoO$_2$ for PLD is the substrate temperature of 620–700°C and the oxygen pressure of 0.1–1.7 Torr.[29, 31] For MBE growth, the substrate temperature ranges in 300–480°C under atomic oxygen plasma[30] or distilled ozone.[32] The room-temperature ab-plane resistivity $\rho_{ab}$ of the PdCoO$_2$ thin films reported to date is about 2-5 times higher than that of the bulk single crystal ($\rho_{ab} = 2.6$ μΩcm).[6, 29] The low-temperature resistivity has been reported for these thin films. The largest residual resistivity ratio (RRR) for PdCoO$_2$ of 16 has been reported in a thick (~180 nm) MBE-grown PdCoO$_2$ thin film that is post-annealed at 800 °C in oxygen.[30] The RRR of 16 in the PdCoO$_2$ thin film is still small compared with the bulk value (RRR~400).[11]

To improve the thin-film quality, dominant scattering sources should be clarified. One of the possible dominant scattering sources is twin boundaries. For c-axis oriented thin-film growth of metallic delafossites, Al$_2$O$_3$ substrates are widely used. As the surface of Al$_2$O$_3$ substrates is effectively 6-fold symmetric, the reported thin films have crystal twins that are 180°-rotated from each other.[29, 30, 32] These twin boundaries can scatter electrons. Growing single crystal thin films without twin boundaries could dramatically improve electron mobility. Although various promising approaches, as listed in Table 1, to achieve this using single crystal substrates with delafossite structures[48] or twin-free delafossite buffer layers[49] have been reported, twin-free thin films have not realized. Toward twin-free thin films, stacking faults need to be also suppressed.

Other possible scattering sources include point defects caused by impurities, cation off stoichiometry, and oxygen vacancies. As the RRR of thin films have not been significantly dependent on the purity of the material sources (PLD targets, MBE sources),[32, 50] impurity scattering seems not a dominant scattering mechanism in the current metallic delafossite thin films. Increase of RRR by post annealing could be due to the reduction of oxygen vacancies as well as improved atomic ordering in the crystal. Post-annealing temperature is limited by the decomposition temperature of metallic delafossites (800-925°C). Alternative approach such as thin-film growth techniques based on chemical reactions could be promising to improve atomic ordering as they may mimic the bulk synthesis of pure single crystals.[33, 51]

As a potentially low-cost and wide-area deposition technique that is important for application, solution-based growth on 2-inch wafers has been reported for PdCoO$_2$, PdCrO$_2$, PtCoO$_2$.[33] Regarding device fabrication, PdCoO$_2$ has been successfully patterned into submicron scales by conventional electron beam lithography and Ar-ion

**Table 1.** Challenges in growth and processing of metallic delafossite thin films and tested approach

| Growth/processing challenges | Tested approach | Applied materials |
|---|---|---|
| Twinning | Miscut substrates[31] | $PdCoO_2$ |
| | Substrates with low symmetry surfaces | |
| |     $\beta$-$Ga_2O_3$ (-201) (ref.[55]) | $PdCoO_2$, $PdCrO_2$ |
| |     $SrTiO_3$ (111) (ref.[49]) | $PdCrO_2$ |
| | Buffer layers: $CuCrO_2$ (ref.[34, 49]) | $PdCrO_2$ |
| | Delafossite-type substrates: $CuFeO_2$ (ref.[48]) | – |
| Stacking faults | Observed by HAADF-STEM[27, 30] | $PdCoO_2$ |
| Impurities | PLD with high purity targets ($PdCoO_2$ 4N5, Pd 5N)[50] | $PdCoO_2$ |
| | MBE with high purity sources (Pd 5N, Co 4N5)[32] | $PdCoO_2$ |
| Cation stoichiometry | Alternate ablation of $PdCoO_2$ and $PdO_x$ in PLD[29] | $PdCoO_2$ |
| | MBE growth[30, 32] | $PdCoO_2$ |
| Oxygen vacancies | Post annealing in air or $O_2$ improves *RRR* (ref.[30]) | $PdCoO_2$ |
| | MBE in atomic oxygen plasma[30] or ozone[32] | $PdCoO_2$ |
| Atomic ordering | Post annealing in air or $O_2$ improves *RRR* (ref.[30]) | $PdCoO_2$ |
| | Chemical growth method | |
| |     Solution-based growth[33] | $PdCoO_2$, $PdCrO_2$, $PtCoO_2$ |
| |     Atomic layer deposition[51] | $PtCoO_2$ |
| Lattice matching (*ab*-plane and *c*-axis) | Buffer layers: $CuCrO_2$ (ref.[34, 49]) | $PdCrO_2$ |
| | Delafossite-type substrates: $CuFeO_2$ (ref.[48]) | – |
| Wide area | Solution-based growth on 2-inch wafers[33] | $PdCoO_2$, $PdCrO_2$, $PtCoO_2$ |
| Heterostructures | Schottky junctions with $\beta$-$Ga_2O_3$ (ref.[35, 55-57]) | $PdCoO_2$, $PdCrO_2$ |
| Microfabrication | Electron beam lithography and Ar ion milling[27] | $PdCoO_2$ |
| | Resist: HSQ, Line width: ~100 nm | |

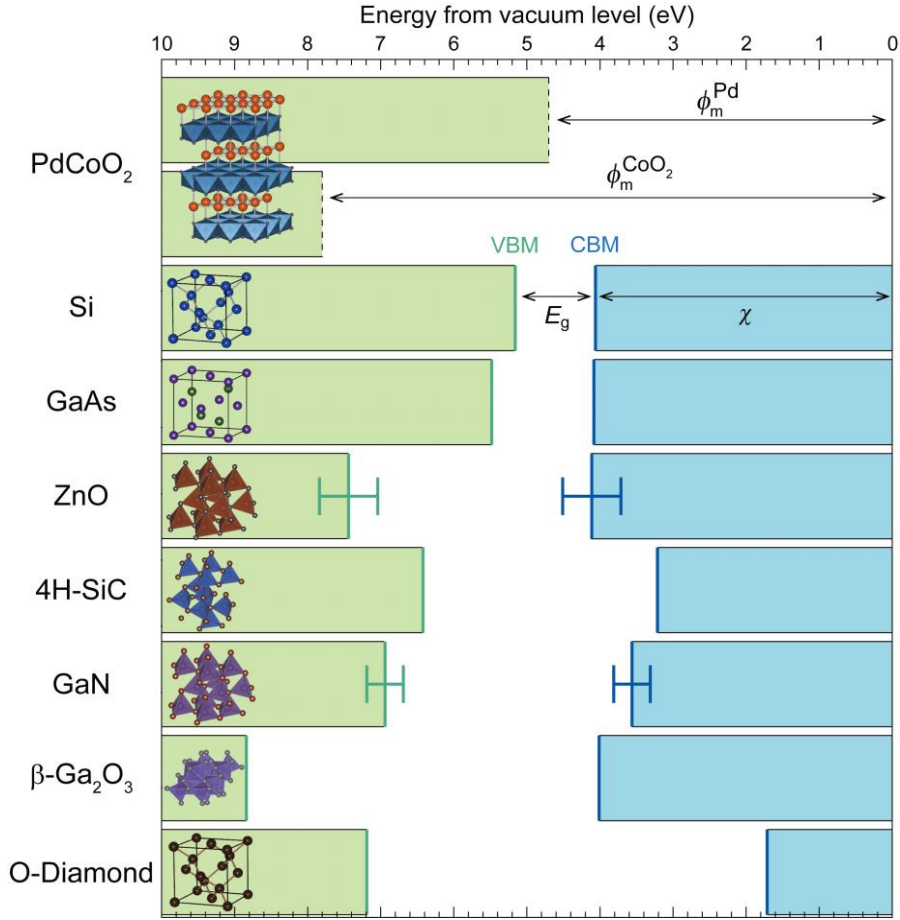

**FIG. 4.** The band alignment of semiconductors compared with the work function of Pd-terminated ($\phi_m^{Pd}$)[53, 55] and CoO$_2$-terminated ($\phi_m^{CoO2}$)[53] PdCoO$_2$. The valence and conduction bands of the various semiconductors are shown in green and blue, respectively. For Si, the electron affinity ($\chi$), the bandgap ($E_g$), the conduction band minimum (CBM), and the valence band maximum (VBM) are noted. The band alignment is drawn based on the reported bandgap and the electron affinity for Si,[93] GaAs,[93] ZnO,[93, 94] 4H-SiC,[95] GaN,[96] $\beta$-Ga$_2$O$_3$,[97, 98] and oxidized diamond surfaces.[99, 100] For ZnO and GaN, the CBM and VBM are shown with bars, which are reported to depends on the surface termination.[94, 96]

milling.[27] As discussed in the sections 5 and 6, search for novel heterostructures is important to find a possible application of metallic delafossites.

## 5. POLAR SURFACES FOR DEVICE APPLICATIONS

Along with high electron mobility/conductivity, the surface polarity of metallic delafossites is also highly significant as it permits an exceptionally large work function, which is beneficial for device applications. According to the classification by Tasker,[52] the $c$-plane surfaces of metallic delafossites are classified as type-3 polar surfaces.[6] As a result, the surface properties of metallic delafossites strongly depend on the termination layer of the surface (e.g., Pd or CoO$_2$ layer). Recent surface measurements of cleaved PdCoO$_2$ showed that the work function of a Pd-terminated surface ($\phi_m^{Pd}$) was ~4.7 eV while a CoO$_2$-terminated surface ($\phi_m^{CoO2}$) was ~7.8 eV, indicating that the work function of PdCoO$_2$ is highly dependent on the termination layer.[53] The large difference between $\phi_m^{Pd}$ and $\phi_m^{CoO2}$ is due to the polar surface. A work function of ~7.8 eV is very large for a metallic compound and far exceeds the Pt work function (~5.65 eV), which is the largest value found in elemental metals.[54]

The large work function of PdCoO$_2$ should be of great value for various device applications, such as for electrodes for semiconductor devices. In Figure 4, the work function of PdCoO$_2$ ($\phi_m^{Pd}$ and $\phi_m^{CoO2}$) is compared with the conduction band minimum (CBM) and valence band maximum (VBM) of various semiconductors. In semiconductor devices, Schottky junctions formed at metal/semiconductor interfaces often play key roles in device operation. According to the Schottky-Mott relationship, the ideal Schottky barrier height of metal/n-type semiconductor interfaces, $\phi_{b,n}$, is estimated to be $\phi_{b,n} = \phi_m - \chi$, where $\phi_m$ is the work function of a metal and $\chi$ is the electron affinity of a semiconductor. As shown in Figure 4, the difference between $\phi_m^{CoO2}$ and $\chi$ of various semiconductors is as large as several eV. Therefore, large Schottky barrier heights can be achieved in PdCoO$_2$/n-type semiconductor polar interfaces, as demonstrated by PdCoO$_2$/$\beta$-Ga$_2$O$_3$ [55-57] and PdCrO$_2$/$\beta$-Ga$_2$O$_3$ junctions.[35] The large Schottky barrier height would be useful for diodes for high-temperature operation and Schottky gates for transistors. Careful control of the initial growth layer would be important to generate an interface with a homogeneous Schottky barrier height.

In general, ohmic contacts with low contact resistance are important to reduce power loss in wide bandgap semiconductor devices. Making low resistance contacts to p-type wide bandgap semiconductors is often a challenge, particularly when the hole concentration is low. In the Schottky-Mott relationship, the Schottky barrier height of a metal/p-type semiconductor interface is $\phi_{b,p} = \chi + E_g - \phi_m$, where $E_g$ is the bandgap of the semiconductor. To achieve a true ohmic contact ($\phi_{b,p} < 0$), one needs a metal with $\phi_m > \chi + E_g$. For wide bandgap semiconductors such as 4H-SiC ($E_g \sim 3.2$ eV) and GaN ($E_g \sim 3.4$ eV), $\chi + E_g$ is well above 6 eV, and there is no elemental metal electrode that satisfies $\phi_m > \chi + E_g$. As shown in Figure 4, $PdCoO_2$ with a $CoO_2$ initial layer is a promising candidate for a true ohmic contact for wide bandgap semiconductors since $\phi_m^{CoO2}$ is larger than the $\chi + E_g$ of semiconductors with p-type doping (Si, GaAs, ZnO, 4H-SiC, GaN, and diamond).

## 6. SURFACE SPIN STATES

Surface polarity also causes emergent spin polarized surface states. In general, to electrostatically stabilize the polar surface, charge compensation occurs.[52, 58, 59] This causes the electron density at the surface to differ from that of the inner bulk. Recent experiments have indicated that charge compensation changes the surface of the inherently nonmagnetic $PdCoO_2$ to have spin-dependent electronic states.

The surface electronic states of $PdCoO_2$ have been calculated using a density functional theory (DFT)-based method.[60] Owing to the electronic reconstruction induced by the surface polarity, a Pd-terminated surface of $PdCoO_2$ was proposed to have surface magnetism.[60] The calculation also predicted that the Co-derived bands at a $CoO_2$-terminated surface are split because of the spin-orbit interaction.[60]

Sunko and colleagues observed the $BO_2$-terminated surfaces of $PdCoO_2$, $PtCoO_2$, and $PdRhO_2$ by angle-resolved photoemission spectroscopy (ARPES) and discovered Rashba-like spin splitting,[22] which is caused by strong inversion symmetry breaking on the $BO_2$-terminated surface. Spin splitting is comparable to atomic spin-orbit coupling of $B$-site cations. Mazzola *et al.* reported the surface ferromagnetic states at Pd-terminated surfaces of $PdCoO_2$ by ARPES.[23] Ferromagnetism has also been detected in PLD-grown $PdCoO_2$ thin films by ARPES and the anomalous Hall effect.[24] From the magnetoresistance measurements under the in-plane magnetic field, the Pd-terminated surface of $PdCoO_2$ also has Rashba-spin splitting with a Rashba coefficient ~ 0.75 ± 0.3 eV.[61] Possible coexistence of ferromagnetism and Rashba spin-orbit coupling will be an interesting subject for future studies in the context of spin-orbitronics.[62]

Here we will discuss the prospects of using the Rashba ferromagnetic surfaces of metallic delafossites. We performed DFT calculations of a Pt-terminated $PtCoO_2$ surface (Fig. 5) and found similar properties to that of $PdCoO_2$. In the nonmagnetic calculation, Rashba-like split bands were observed around the Γ point. Furthermore, the flat region of the conduction band shifted close to the Fermi level, which may cause Stoner splitting, as with $PdCoO_2$.[23] In the calculation considering collinear spin polarization, the bands around the Γ point were spin-split. The Rashba effect visible in the nonmagnetic calculation was more pronounced in $PtCoO_2$ compared with that in $PdCoO_2$ because of the larger atomic spin-orbit coupling in Pt than in Pd (Fig. 5(a)).

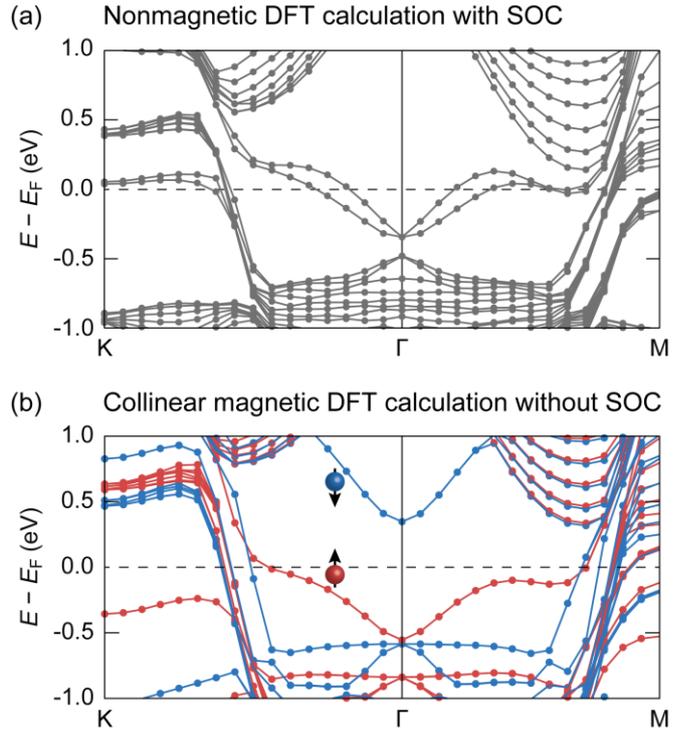

**FIG. 5.** The electronic structure of the slab model for a Pt-terminated $PtCoO_2$, calculated by density functional theory (DFT) using the CASTEP package. The vertical axis is the energy relative to the Fermi level ($E_F$). (a) The nonmagnetic calculation result considering spin-orbit coupling (SOC). (b) The calculation result considering collinear spin-polarization. The red and blue lines correspond to the spin-up and spin-down states, respectively. For (b), SOC is not included because of the limitation of our computer setup. Including spin-orbit interaction can give rise to non-collinear spin textures and require larger supercells for calculation. A symmetric slab model containing nine Pt and eight $CoO_2$ layers with a vacuum gap of 15 Å is employed for the calculation, following a report on $PdCoO_2$.[23] The inner five Pt and four $CoO_2$ layers are fixed to have the lattice constants of bulk single crystals.[4] The outer four Pt and four $CoO_2$ layers were relaxed. For both (a) and (b), the generalized gradient approximation and Perdew-Burke-Ernzerhof (GGA-PBE) functionals are used, and the Hubbard $U$ is not taken into account.

Surface spin-split electronic states can be utilized in heterostructures to induce exotic electronic states. For example, making a heterostructure with a superconductor may be interesting. As a model system that could show topological superconductivity, a 2D electronic system with relatively large spin-orbit coupling, effective Zeeman coupling, and proximity-induced superconductivity has been proposed.[63, 64] The model Hamiltonian can be written as $H(k) = H_0(k) + H_R(k) + H_Z + H_S$, where $H_0$ is kinetic energy, $H_R$ is the Rashba spin-orbit interaction, $H_Z$ is the effective Zeeman interaction, and $H_S$ is the spin-singlet s-wave pair potential.[64] At the surface of $PtCoO_2$ or $PdCoO_2$, there can be a relatively large spin-orbit coupling $H_R(k)$ and

Stoner splitting (i.e., effective Zeeman coupling, $H_Z$) in a nearly 2D conduction channel. Inducing superconductivity $H_S$ by the proximity effect and applying a magnetic field to adjust $H_Z$ might induce topological superconducting states.

## 7. REAL SPACE PROBING OF ELECTRONIC STATES

To search for exotic electronic states in metallic delafossite, it is important to understand the external magnetic field dependence of electronic states. An ideal experimental approach for this purpose is SI-STM, which has a unique capability to resolve differential conductance ($dI/dV(r,E,B)$) that is proportional to the density of state bellow and above $E_F$, as a function of position $r$ with atomic precision, energy $E$, and magnetic field $B$.[65-70] SI-STM has been applied to the polar surfaces of $PdCoO_2$ bulk single crystals. On $CoO_2$-terminated surfaces of $PdCoO_2$, long-lived quasiparticle states have been visualized as extended six-fold standing waves.[53] This observation is consistent with the anisotropic hexagonal cylindrical Fermi surface of high mobility electrons (Fig. 3(b)). The observed standing waves have been interpreted by considering the large Rashba-like band splitting driven by strong inversion symmetry breaking.[53] On Pd-terminated surfaces of $PdCoO_2$, SI-STM has revealed the formation of charge density waves[71] and renormalized electronic states that are coupled with magnons.[72]

To further understand the electronic states of $PdCoO_2$ surfaces, SI-STM under external magnetic fields would be informative. As discussed in section 6, metallic delafossites are expected to contain characteristic surface spin states. On Pd-terminated surfaces of $PdCoO_2$ (and possibly in $PtCoO_2$), the possible coexistence of $H_R$ and $H_Z$ may give rise to spin-momentum coupled electronic structures with a $H_Z$-derived gap at the band crossing point (Fig. 5). Such spin-momentum coupled band structures with surface ferromagnetism could respond to external magnetic fields with controlled magnitude and direction.[73,74] For example, studies on magnetic-field dependent standing wave patterns and spatially averaged $dI/dV(E)$ shapes could reveal the spin-momentum coupled states of $PdCoO_2$. As SI-STM can visualize spin texture with atomic resolution in real space[75-78], searching for topological magnetism on the surface of metallic delafossites is an exciting prospect.

To the best of our knowledge, all SI-STM studies reported to date were performed on cleaved surfaces of single crystals. By applying SI-STM to metallic delafossite thin films (Fig. 6), we would be able to investigate the electronic states of heterostructures with other materials, such as the superconductors discussed in section 6. Furthermore, as with other layered materials, monolayer-thick metallic delafossites may also be an interesting system for SI-STM study.

Finally, another important task is to identify the scattering sources that limit electron mobility of thin films. The electronic scattering response to twin-domain boundaries and point defects could be visualized by SI-STM, which would provide feedback for thin-film growth optimization. The combination of thin-film growth and real-space imaging techniques may lead to high-mobility thin films, facilitating research into unexplored phenomena in ultrathin films and heterostructures.

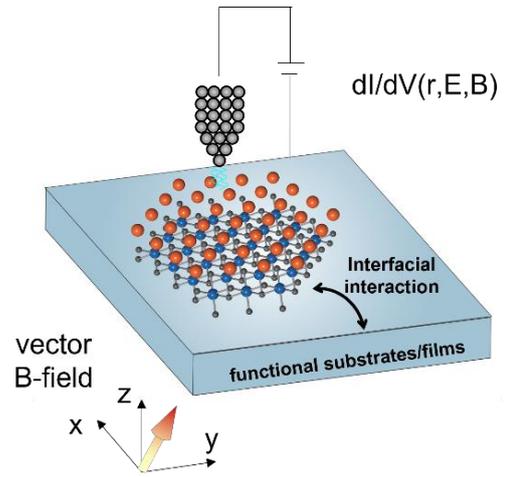

**FIG. 6.** A conceptual figure for spectroscopic imaging scanning tunneling microscopy (SI-STM) on metallic delafossite thin films and heterostructures. SI-STM is able to visualize electronic states with ultra-high spatial and energy resolution under controlled external magnetic fields.

## 8. CONCLUSION

In this article, we have provided an overview of the physical properties of metallic delafossites and discussed the perspectives of thin-film research. Inherent high mobility and surface polarity make metallic delafossites fascinating materials for both basic and application studies. In particular, mesoscopic electrical transport, many-body effects derived from Fermiology, and the interplay of these may lead to intriguing unexplored phenomena. Thin film growth of metallic delafossites would provide tremendous research possibilities for the design of heterostructures and functional devices.

**Acknowledgments**

The author would like to thank Prof. A. Tsukazaki and all collaborators for their contributions to this work on metallic delafossites. This work was supported by a Grant-in-Aid for Scientific Research (B) (20H02611), a Grant-in-Aid for Challenging Research (exploratory) (20K21138) from the Japan Society for the Promotion of Science (JSPS), MEXT Leading Initiative for Excellent Young Researchers (JPMXS0320200047), JST PRESTO (JPMJPR20AD), Izumi Science and Technology Foundation, and the Foundation for The Promotion of Ion Engineering.